# Title: Direct observation of 2D magnons in atomically thin $CrI_3$

**Authors:** John Cenker[1†], Bevin Huang[1†], Nishchay Suri[2], Pearl Thijssen[1], Aaron Miller[1], Tiancheng Song[1], Takashi Taniguchi[3], Kenji Watanabe[3], Michael A. McGuire[4], Di Xiao[2]*, Xiaodong Xu[1,5]*

[1]Department of Physics, University of Washington, Seattle, Washington 98195, USA
[2]Department of Physics, Carnegie Mellon University, Pittsburgh, Pennsylvania 15213, USA
[3]National Institute for Materials Science, 1-1 Namiki, Tsukuba 305-0044, Japan
[4]Materials Science and Technology Division, Oak Ridge National Laboratory, Oak Ridge, Tennessee 37831, USA
[5]Department of Materials Science and Engineering, University of Washington, Seattle, Washington 98195, USA

[†]These authors contributed equally to this work.

*Corresponding author's e-mail: xuxd@uw.edu; dixiao@cmu.edu

**Abstract: Exfoliated chromium triiodide ($CrI_3$) is a layered van der Waals (vdW) magnetic insulator that consists of ferromagnetic layers coupled through antiferromagnetic interlayer exchange[1–3]. The resulting permutations of magnetic configurations combined with the underlying crystal symmetry produces tunable magneto-optical phenomena that is unique to the two-dimensional (2D) limit[4–7]. Here, we report the direct observation of 2D magnons through magneto-Raman spectroscopy with optical selection rules that are strictly determined by the honeycomb lattice and magnetic states of atomically thin $CrI_3$. In monolayers, we observe an acoustic magnon mode of ~0.3 meV with cross-circularly polarized selection rules locked to the magnetization direction. These unique selection rules arise from the discrete conservation of angular momentum of photons and magnons dictated by threefold rotational symmetry in a rotational analogue to Umklapp scattering[8]. In bilayers, by tuning between the layered antiferromagnetic and ferromagnetic-like states, we observe the switching of two magnon modes. The bilayer structure also enables Raman activity from the optical magnon mode at ~17 meV (~4.2 THz) that is otherwise Raman-silent in the monolayer. From these measurements, we quantitatively extract the spin wave gap, magnetic anisotropy, intralayer and interlayer exchange constants, and establish 2D magnets as a new system for exploring magnon physics.**

## Main Text:

Strong magnetic anisotropy in a two-dimensional (2D) spin system leads to stable long-range magnetic order[1,9,10]. A direct consequence is the formation of a sizeable spin wave gap, which may lead to magnons with sufficient energies to be probed by Raman spectroscopy[9,11]. This could enable the study of weak magnon features even in the atomically thin limit due to the higher signal-to-noise ratio that Raman scattering offers as compared to Brillouin light scattering (BLS), a spectroscopy technique often used to detect gigahertz magnons in ferromagnets[12]. One such candidate material with strong anisotropy is the ferromagnetic (FM) insulator $CrI_3$, which is theoretically predicted to have a spin wave gap of 0.3 to 0.4 meV (2.4 – 3.2 $cm^{-1}$)[11,13] in the monolayer. In addition to these theoretical predictions, Raman spectroscopy studies have revealed spin wave features in samples as thin as ten layers[5].

Layered 2D magnets also possess unique properties for investigating magnon physics[14–16]. For instance, $CrI_3$ monolayers have a honeycomb lattice structure belonging to the $D_{3d}$ point group, thereby offering a pristine crystal lattice with threefold rotational symmetry. Distinct from systems with continuous rotational symmetry, the honeycomb lattice should enable magnon optical selection rules in the backscattering geometry[8,17], which is necessary for 2D crystals. With two $Cr^{3+}$ ions per unit cell, two branches of spin waves are expected corresponding to an acoustic mode with in-phase precession (Fig. 1a) and an optical mode with out-of-phase precession (Fig. 1b)[18]. In bilayer $CrI_3$, the individual FM monolayers become antiferromagnetically (AFM) coupled[1], leading to a layered AFM ground state. When a magnetic field of ~0.7 T is applied, a metamagnetic transition occurs that leads to an FM-like state. Switching between these magnetic states provides direct control over the symmetry of the bilayer, leading to giant second-harmonic generation[4] and tuning of inelastic scattering from spin-coupled phonons[6,7]. Furthermore, the ability to externally tune the magnetic order through mechanisms such as pressure[19,20] and electrostatic gating[21,22] opens opportunities for novel spintronic and magnonic devices. The combination of symmetry and magnetism thus could provide a rich platform to study spin wave physics in the atomically thin limit.

In our experiments, we sandwiched exfoliated monolayer and bilayer $CrI_3$ between 20-30 nm flakes of hexagonal boron nitride (hBN). Magneto-Raman measurements were performed using a 632.8 nm HeNe laser (see Methods), which is nearly resonant with a ligand-to-metal charge transfer transition at 2 eV,[23] in the backscattering geometry. That is, both the incident and scattered light are normal to the sample plane. The magnetic behavior of these samples was confirmed through polar reflective magnetic circular dichroism (RMCD) measurements in the same setup. A representative $CrI_3$ monolayer is shown in Fig. 1c (optical micrograph, left inset). The zero-field RMCD mapping and RMCD sweep (right inset) confirms that the flake is a hard ferromagnet with a single domain.

Figure 1d shows the cross-circularly polarized low-frequency Raman spectra at select magnetic fields (see Figure S1 for full Raman spectra including co-circular channels). To suppress the strong laser scattering, we used an optical filter with a cutoff near ±3 $cm^{-1}$ (see Methods), indicated by the gray shaded box. At zero field, the spectra are dominated by leakage of the Rayleigh scattered light. When a magnetic field is applied, a low-frequency feature emerges in only the cross-circularly polarized channel (Figs. 1e&f, see Fig. S1 for co-polarized spectra). The low-frequency mode appears on both Stokes and anti-Stokes sides, which are labeled as AM and AM*, respectively. These Stokes and anti-Stokes modes obey distinctly opposite optical selection rules: for a given magnetization orientation, AM only appears when excited by one helicity of light, while AM* emerges with excitation of the opposite helicity. For instance, with the magnetization pointing down (Fig. 1e), only $\sigma^+$ excitation can create the AM mode with nearly perfect $\sigma^-$ polarization ($\sigma^+/\sigma^-$), while the $\sigma^+$ polarized AM* mode is excited by $\sigma^-$ polarized light ($\sigma^-/\sigma^+$). When the direction of the magnetization is flipped, the selection rules correspondingly reverse, as seen in Fig. 1f.

Figure 2a shows the intensity plot of AM and AM* as a function of Raman shift and magnetic field. The extracted peak positions at each field are plotted in Fig. 2b. Both Stokes and anti-Stokes peaks shift linearly with applied field at a slope of 0.94 $cm^{-1}$/T. This is consistent with the Zeeman energy shift of a spin-1 quasiparticle with a magnetic moment of 2 $\mu_B$, the signature of a magnon. Extrapolating the magnon energy to zero field yields the spin wave gap, which we determine to be ~2.4 $cm^{-1}$ (or ~0.3 meV) in monolayer $CrI_3$. This spin wave gap is quite large for a ferromagnet due to the large magnetic anisotropy and is in excellent agreement with the theoretical predictions in refs. [11] and [13]. We note that the monolayer spin wave gap is also consistent with the results from Raman measurements on bulk single crystal $CrI_3$ (see Figure S2). This implies that the acoustic magnon properties are determined mainly by out-of-plane magnetic anisotropy.

Temperature-dependent measurements further confirm the magnetic origin of AM and AM*. Figure 2c shows the AM and AM$^*$ intensity plotted as a function of temperature and Raman shift. The applied magnetic field is kept fixed at -7 T, thereby shifting the magnon mode away from the Rayleigh scatter and stabilizing the spin-polarized state at higher temperatures. This allows for the study of the acoustic magnon through a wide temperature range. The magnon feature persists and slightly redshifts until temperatures as high as 80 K at which point the scattering becomes too weak to detect. It is worth noting that the linewidth remains essentially constant through the whole temperature range (see Fig. S3 for Raman spectra at select temperatures). This implies that the true linewidth, $\Gamma$, is limited by the resolution of our spectrometer. As a result, we infer a lower bound of the lifetime, $\tau$, as $\Gamma^{-1}$ = 5 ps. These results call for further ultrafast time-resolved measurements in order to determine the true magnon lifetime and its dependence on temperature and magnetic field.

The distinct cross-circular optical selection rules can be derived by considering the angular momentum of the photon, $J_p$, magnon, $J_m$ in a honeycomb lattice (Fig. 2d). For instance, when the magnetization points up, the total change of photon angular momentum is $\Delta J_p = \pm 2\hbar$ since the helicities between incident and scattered light are opposite, where + (−) corresponds to anti-Stokes (Stokes) scattering from the magnon mode. The change of magnon angular momentum contributes $\Delta J_m = \pm\hbar$. This leads to a total change of angular momentum that is $\Delta J_p + \Delta J_m = \pm 3\hbar$. In a system with continuous rotational symmetry, the conservation of angular momentum would forbid such a process. However, in an analogue to the Umklapp process[8,17], the threefold symmetry of the honeycomb lattice in monolayer $CrI_3$ allows for discrete angular momentum conservation up to modulo $3\hbar$: $|\Delta J_p + \Delta J_m|/\hbar = 0 \pmod 3$. Given that the total change of the angular momentum is a multiple of $3\hbar$, cross-circular selection rules are expected in monolayer $CrI_3$. A group theory analysis yields the same conclusion (see Methods). We note that for bulk single crystals (Figure S2), we observe a relaxation of this selection rule. The magnon scattering is still predominately in the cross-circular channels, but significant scattering in the co-circular channels is also present. This deviation from the expected selection rules could potentially originate from stacking faults and highlights the intriguing physics afforded by studying atomically thin samples with pristine crystal structures.

Unlike the monolayer, bilayer $CrI_3$ hosts either a ferromagnetic-like state or a layered antiferromagnetic state, depending on the applied magnetic field, below a critical temperature of 45 K. Figure 3a shows the low-frequency Raman spectrum of a $CrI_3$ bilayer in an applied field of 6 T. Since the magnetization of both layers are aligned in the same direction by the magnetic field, the bilayer essentially behaves like a ferromagnetic monolayer. That is, the acoustic magnon mode is cross-circularly polarized with the Stokes and anti-Stokes peaks showing opposite selection rules and shifts linearly in energy with a $g$ factor of ~2.1 under applied field (Fig. 3b). However, in contrast to the monolayer, we observe switching of acoustic magnon modes as the bilayer undergoes a metamagnetic transition between -0.7 T and -0.75 T (see Fig. S4 for RMCD measurements with consistent transition field). Figure 3c shows a series of Raman spectra as the magnetic states switch from the AFM to the FM-like state. Since the shift of the magnon peak over small field ranges near the metamagnetic transition is smaller than the resolution of our spectrometer, we utilize Lorentzian fits to track subtle changes in the magnon energy (see Fig. S5). In the AFM state, at fields close to the metamagnetic transition, the acoustic magnon mode has a frequency of ~3.6 cm$^{-1}$. As the field passes the metamagnetic transition, the magnon mode at 3.6 cm$^{-1}$ vanishes. Instead, another acoustic mode, starting at a lower frequency of ~3.0 cm$^{-1}$, appears and linearly shifts to higher energy as the magnitude of the field increases.

The zero-field magnon energy of the antiferromagnetic acoustic magnon mode, which we find to be ~0.37 meV from a linear fit (green line in Fig. 3b), is given by $\sqrt{(2J_{\text{inter}} + K)K}$, where $J_{\text{inter}}$ is the antiferromagnetic interlayer exchange interaction, and $K$ is the magnetic anisotropy[24]. We can extract the magnetic anisotropy by extrapolating the linear fit (blue line, Fig. 3b) of the acoustic magnon energy shift in the FM-like state to zero field. Doing so results in $K$ ~ 0.27 meV which allows us to find $J_{\text{inter}}$ ~ 0.11 meV. Therefore, bilayer $CrI_3$ is in the weak exchange limit ($K > J_{\text{inter}}$). As the magnetic field increases, instead of a spin-flop transition, bilayer $CrI_3$ undergoes an abrupt spin-flip transition[1], explaining the observed discontinuity of the magnon frequency at the spin-flip field. The weak interlayer exchange $J_{\text{inter}}$ also explains why the cross-circular optical selection rules continue to hold for bilayer $CrI_3$ even though the crystal structure lacks three-fold rotational symmetry due to its monoclinic stacking[4,25–28].

In addition to the low-frequency acoustic magnons, we also resolve a high-frequency cross-polarized mode in bilayers which has not been observed in previous Raman studies on $CrI_3$. Figure 4a shows a very weak and broad peak centered at ~148 cm$^{-1}$ (or 4.4 THz) at 8 T which exhibits the same helicity-dependent selection rules as the acoustic magnon mode. Magnetic field-dependent measurements show that the peak position shifts linearly with a $g$ factor of ~2 (Fig. 4b). We thus assign this peak as the high-frequency optical magnon (OM). We note that the optical magnon mode does not show up in monolayers because it has the form $S_A^- - S_B^-$, where $S_A^-$ and $S_B^-$ are the spin wave basis on the two different sublattices in the unit cell. This excitation is thus parity-odd and Raman-silent. In bilayers, the optical modes from both layers will hybridize, leading to a Davydov-like splitting, with one mode being parity-odd and the other parity-even and consequently Raman-active (see Methods). This optical magnon mode is related to the intralayer exchange by $E = K + 6J_{\text{intra}}$. Therefore, we determine the intralayer exchange to be ~2.83 meV. We also observed the terahertz optical magnon in both exfoliated thin bulk and unexfoliated single crystal samples (see Fig. S6) with frequencies of 160 cm$^{-1}$ (4.8 THz) and 148 cm$^{-1}$ (4.4 THz) respectively. We note that the optical magnon energy is consistent with that seen in recent neutron scattering experiments[18].

Unlike the low-frequency acoustic magnon which merges with Rayleigh scatter, we are able to resolve the OM mode down to zero applied field and observe its behavior in the AFM state. In stark contrast to the FM-like state, the OM mode in the AFM state can be excited by both helicities of incident light, giving equal scattering intensity in both $\sigma^+/\sigma^-$ and $\sigma^-/\sigma^+$ detection channels in the absence of applied field (Fig. 4c). Applying a small field of 0.5 T, however, results in an energy splitting between the OM features seen in the two cross-circular channels shown in Fig. 4d: the OM mode in the $\sigma^-/\sigma^+$ channel blueshifts while the peak in the $\sigma^+/\sigma^-$ channel redshifts. Flipping the direction of the applied field to -0.5 T in Fig. 4e reverses the splitting such that the OM mode in the $\sigma^+/\sigma^-$ ($\sigma^-/\sigma^+$) channel is at a higher (lower) frequency.

In general, antiferromagnets with easy axis anisotropy, such as bilayer $CrI_3$, host two magnon modes per branch which are degenerate at zero-field but carry opposite angular momentum[24,29]. The layered antiferromagnetic order breaks inversion symmetry, eliminating the parity criterion and hence allowing Raman activity of both optical magnon modes. Upon the application of a magnetic field, the two modes shift oppositely, *i.e.* one mode will blueshift and the other will redshift. This explains the Raman activity of the optical branch magnons in both cross-circular channels and the opposite energy splitting of the two OM modes in Figs. 4d&e. The acoustic branch magnons also split in an applied magnetic field, but the redshifted magnon mode (denoted by the orange line in Fig. 3b) is overshadowed by Rayleigh scatter, making it unresolvable in our measurements.

In conclusion, we have identified acoustic and optical magnons in atomically thin $CrI_3$ which obey selection rules dictated by discrete angular momentum conservation in a honeycomb lattice. Using the energies of the acoustic magnons, we calculate the strength of anisotropy and interlayer exchange. The optical magnon, on the other hand, allows for the determination of intralayer exchange. Furthermore, in contrast to the low-frequency acoustic magnons and those observed in standard FM spintronic systems, the frequency of the optical magnon is well into the terahertz regime and is comparable to those found in AFM systems[30]. As $CrI_3$ is an insulator, the lifetime of terahertz magnons may be much longer than those in metallic systems[31,32]. Indeed, we determine that the lifetime is on the order of picoseconds, an order of magnitude longer than those seen in recent high-quality metallic FM thin films[32,33]. Our results establish $CrI_3$ as a desirable candidate for studying fundamental magnon physics with symmetry control and exploring magnonic devices in both the gigahertz and terahertz regime.

**Note:** We note that while preparing this manuscript, we became aware of similar work on magnons in few-layer $CrI_3$[35].

**Methods**

**Sample Preparation:**

Bulk $CrI_3$ crystals were grown by direct vapor transport and exfoliated onto 90 nm $SiO_2$/Si substrates. Using the optical contrast between the flakes and the substrate, monolayer and bilayer flakes of $CrI_3$ were identified and encapsulated between two 20-30 nm thick hexagonal boron nitride (hBN) flakes. All steps of the fabrication process were performed in a glovebox with $N_2$ atmosphere.

Encapsulated samples were assembled through a dry transfer technique with a stamp consisting of a poly(bisphenol A carbonate) (PC) film stretched over a polydimethylsiloxane (PDMS) cylinder[34]. The flakes were picked up in the following order before being deposited onto the $SiO_2$/Si substrate: top hBN, $CrI_3$, bottom hBN.

**Optical Measurements:**

RMCD measurements of $CrI_3$ samples were performed in a cold finger cryostat capable of reaching temperatures down to 15 K and applying magnetic fields up to 7 T. Raman measurements of the acoustic magnon mode and of the optical magnon mode in single crystal bulk and exfoliated thin bulk $CrI_3$ were performed in the same setup. Raman measurements of the optical magnon mode in bilayer $CrI_3$ were taken in an attoDry 2100 cryostat utilizing helium as the exchange gas, allowing the sample to be cooled to 1.6 K and the application of magnetic fields up to 9 T.

For both RMCD and Raman measurements, an objective lens was used to focus 632.8 nm light from a HeNe laser to a beam spot of ~1.5 μm in the backscattering geometry. Various laser powers were used for the different samples in order to prevent degradation: monolayer and bilayer measurements of the acoustic magnon utilized 400 μW of power with an integration time of two minutes, bilayer and single crystal bulk measurements of the optical magnon used 1 mW of power and five-minute integrations, and measurements of exfoliated thin bulk $CrI_3$ used 1.5 mW of power with one-minute integrations. The scattered light was collected and dispersed by either an 1800 or 1200 $mm^{-1}$ groove density grating for low-frequency and high-frequency measurements respectively. BragGrate™ notch filters were utilized to reject Rayleigh scattering and enable the study of Raman features down to ~ ±3 $cm^{-1}$ where + (-) corresponds to the Stokes (anti-Stokes) side.

**Symmetry analysis of the Raman tensor:**

In monolayer $CrI_3$, the $Cr^{3+}$ ions form a honeycomb structure with two inequivalent sites (denoted by $A$ and $B$) per unit cell. Consequently, the magnon spectrum has two branches, which we call the acoustic and optical branches following the convention for phonons. The magnetic point group of monolayer $CrI_3$ is $D_{3d}(C_{3i}) = C_{3i} + \theta c_{2x} C_{3i}$, where $\theta$ is the time reversal operator. In the Hermann–Mauguin notation, the magnetic point group is denoted by $\bar{3}m'$. Since the system has inversion symmetry, the magnon modes at the Γ point are eigenmodes of the parity operator. Specifically, the acoustic mode is given by $S_a = S_A + S_B$ and the optical mode by $S_o = S_A - S_B$. The inversion operation switches the $A$ and $B$ sites but leaves spin invariant. Therefore, the optical mode ($S_o$) is parity-odd and hence Raman-silent. On the other hand, the acoustic mode ($S_a$) is parity-even and Raman-active. To determine the Raman tensor for the acoustic mode, we note that $S_a^x$ and $S_a^y$ transform according to the $E_g$ irreducible representation of $\bar{3}m'$. Hence the polarizability tensor has the following expansion in terms of the spin wave basis:

$$\alpha = \begin{pmatrix} iA & -A & B \\ -A & -iA & -iB \\ B & -iB & 0 \end{pmatrix} S_a^+ + \begin{pmatrix} iC & C & D \\ C & -iC & iD \\ D & iD & 0 \end{pmatrix} S_a^-,$$

where $S_a^{\pm} = S_a^x \pm iS_a^y$. These two terms are the contributions from the anti-Stokes and Stokes modes. One can see that these two modes satisfy opposite cross-circularly polarized optical selection rules.

We now consider the FM-like state of bilayer $CrI_3$. In this case, inversion remains a symmetry of the system and we can continue using parity to classify the magnon modes. The magnon modes from the top and bottom layer will hybridize due to the interlayer exchange, giving rise to a Davydov-like splitting. The two optical modes are given by $S_o^- = (S_{A1}^- - S_{B1}^-) \pm (S_{A2}^- - S_{B2}^-)$, where $A1$ denotes the $A$ site in layer 1. One of the modes is parity-odd and thus Raman-silent, while the other is parity-even and Raman-active.

**Acknowledgements:** We thank Nathan Wilson for the helpful discussion. This work was mainly supported by the Department of Energy, Basic Energy Sciences, Materials Sciences and Engineering Division (DE-SC0012509). Device fabrication and understanding of magnon optical selection rules are partially supported by AFOSR MURI 2D MAGIC (FA9550-19-1-0390). Work at ORNL (MAM) was supported by the US Department of Energy, Office of Science, Basic Energy Sciences, Materials Sciences and Engineering Division. K.W. and T.T. acknowledge support from the Elemental Strategy Initiative conducted by the MEXT, Japan and the CREST (JPMJCR15F3),

JST. BH acknowledges partial support from NW IMPACT. XX acknowledges the support from the State of Washington funded Clean Energy Institute and from the Boeing Distinguished Professorship in Physics.

**Author contributions:** XX, JC and BH conceived the experiment. JC and BH fabricated and characterized the samples, assisted by AM and PT. JC and BH performed the Raman and magnetic circular dichroism measurements, assisted by PT and TS. JC, BH, NS, DX, and XX analysed and interpreted the results. TT and KW synthesized the hBN crystals. MAM synthesized and characterized the bulk $CrI_3$ crystals. JC, BH, XX and DX wrote the paper with input from all authors. All authors discussed the results.

**Competing Interests:** The authors declare no competing financial interests.

**Data Availability:** The datasets generated during and/or analysed during this study are available from the corresponding author upon reasonable request.

**Figures:**

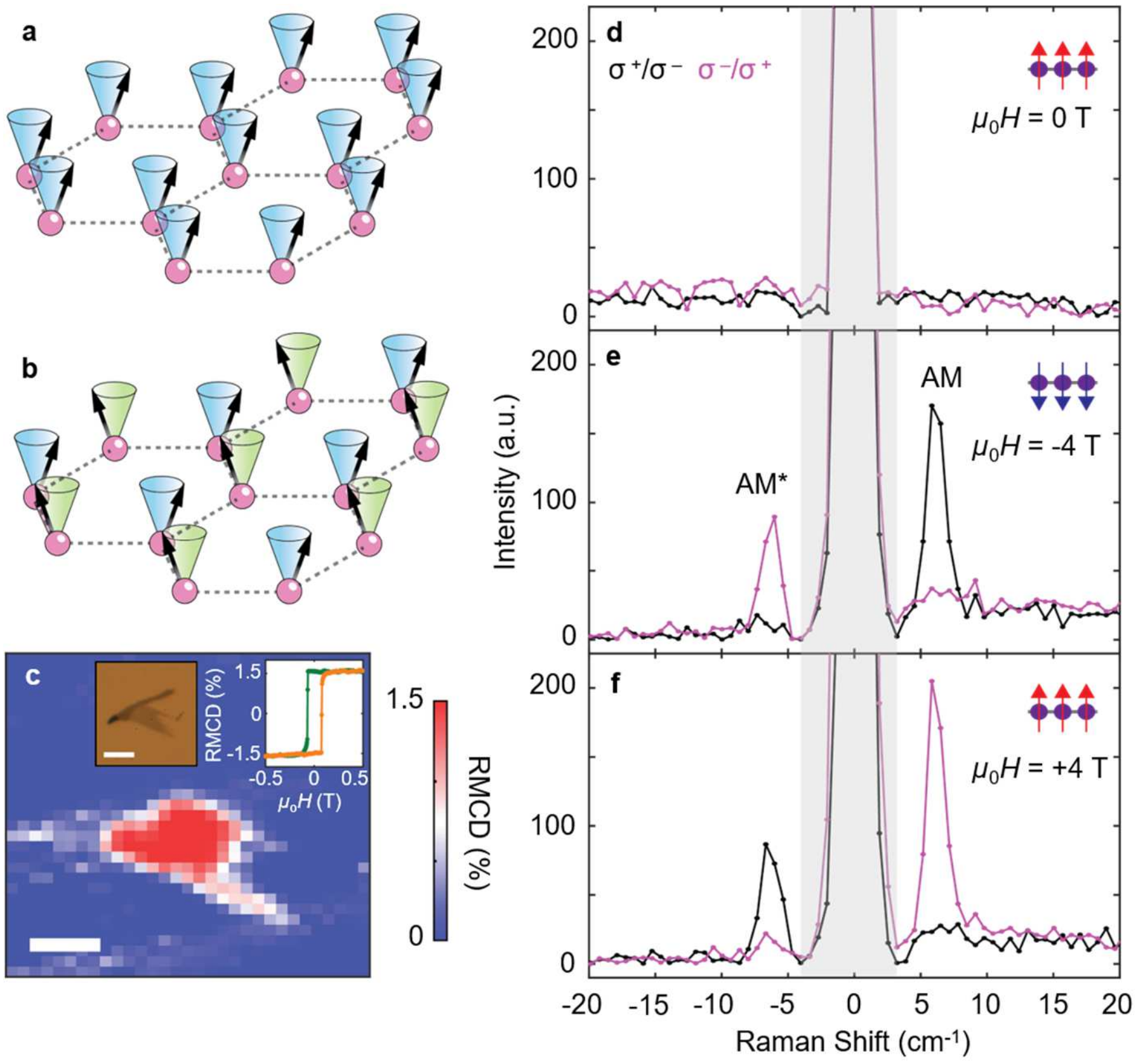

**Figure 1 | Spin wave diagram of monolayer $CrI_3$, sample characterization and low-frequency Raman spectra. a-b,** Depiction of the two types of spin waves in ferromagnetic monolayer $CrI_3$: **a** corresponds to the in-phase acoustic mode while **b** corresponds to the out-of-phase optical one. **c,** Zero-field reflective magnetic circular dichroism (RMCD) map of an encapsulated monolayer $CrI_3$ flake. Scale bar: 4 µm. The right inset shows an RMCD sweep taken at the center of the flake while the left inset shows an optical micrograph of the sample with a scale bar of 10 µm. **d-f,** Low-frequency cross-polarized Raman spectra taken at **d,** zero-field **e,** -4 T, and **f,** 4 T. The grey box indicates the spectral range of our optical filter below which the magnon is unresolvable.

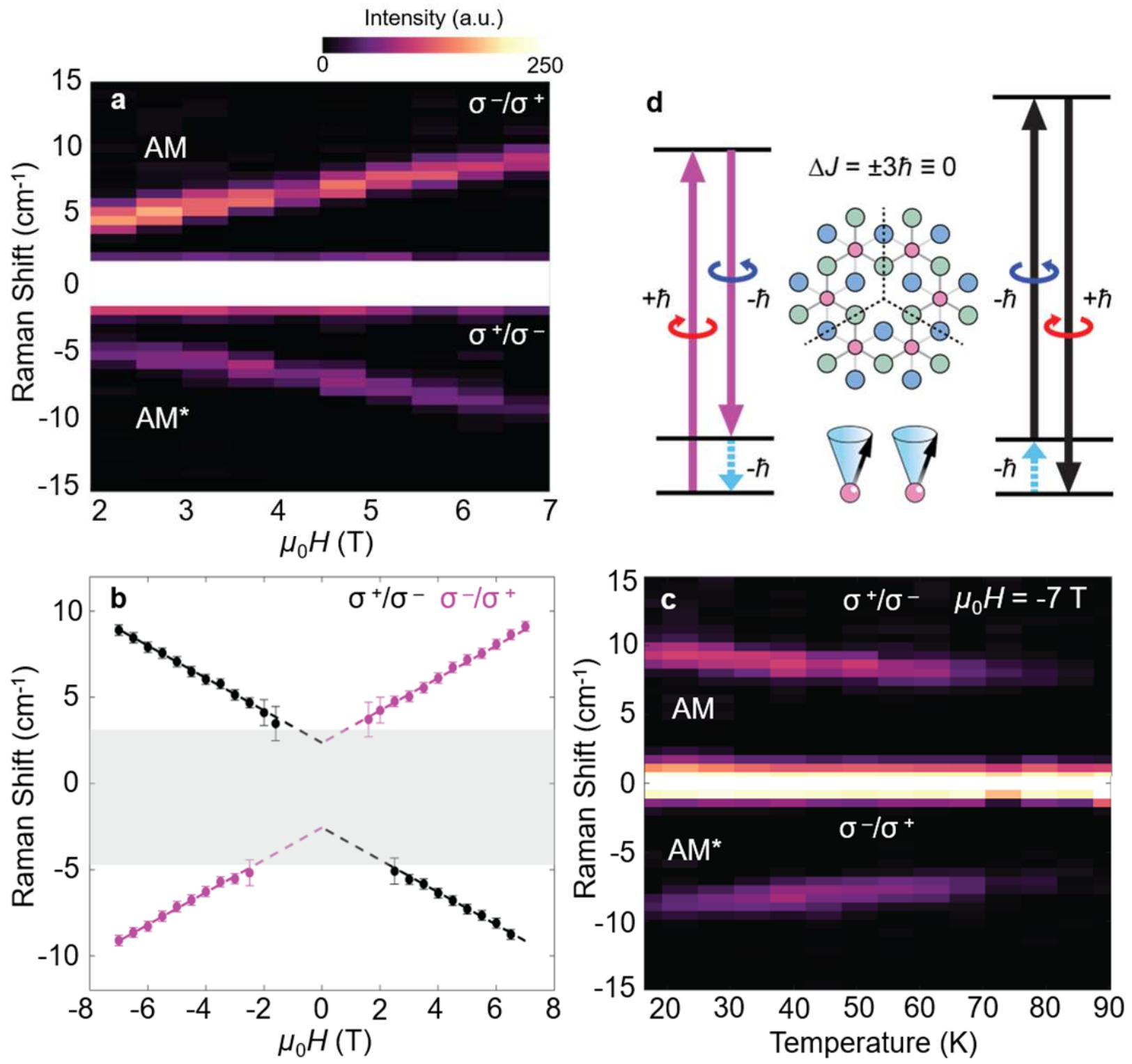

**Figure 2 | Magnetic field and temperature-dependence of spin waves in monolayer $CrI_3$.** **a,** Color map of magnetic field-dependent Raman measurements of monolayer $CrI_3$ in the field range from 2 to 7 T. Data above 0 $cm^{-1}$ is taken in the $\sigma^-/\sigma^+$ channel while data below 0 $cm^{-1}$ is taken in the $\sigma^+/\sigma^-$ channel. **b,** The center of the magnon features obtained by Lorentzian fits of the spectra plotted as a function of field from –7 to 7 T. The dashed line shows the expected Zeeman energy shift for an $S = 1$ quasiparticle with a zero-field energy of 2.4 $cm^{-1}$ (~0.3 meV). The error bars represent the uncertainty of the Lorentzian fit used to determine the peak center. **c,** Color map of temperature-dependent Raman measurements indicate a gradual redshifting with temperature until the scattering vanishes at temperatures above ~80 K. **d,** Optical selection rules of the one-magnon scattering in monolayer $CrI_3$. The left (right) energy diagram corresponds to Stokes (anti-Stokes) scattering in the spin-up state. The total change in angular momentum of $\pm 3\hbar$ is equivalent to 0 in the honeycomb lattice shown in the middle. Pink, green, and blue circles in the honeycomb lattice represent $Cr^{3+}$, top $I^-$, and bottom $I^-$ ions respectively.

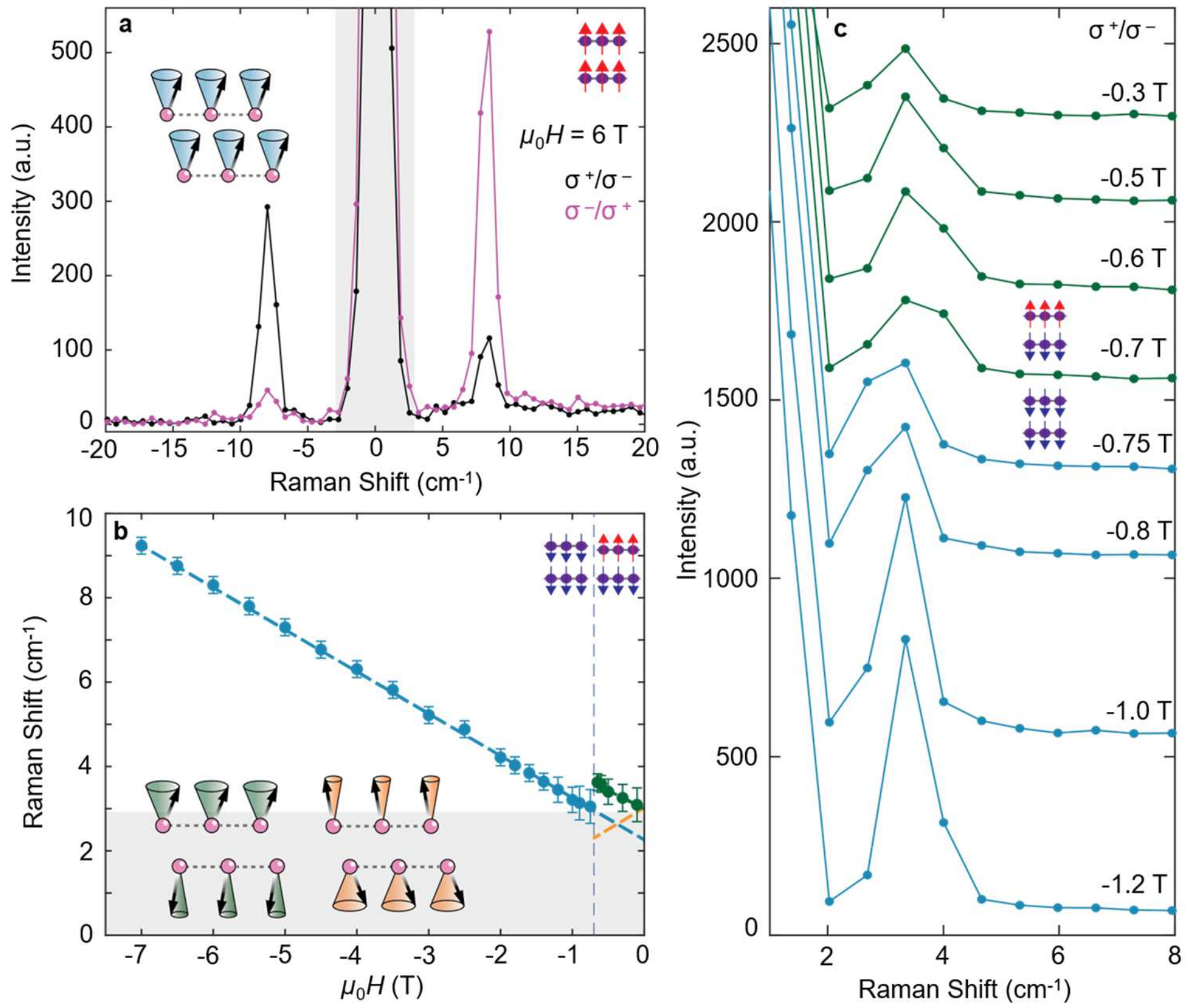

**Figure 3 | Magnon scattering in magnetic $CrI_3$ bilayers. a,** Low-frequency Raman spectrum of a $CrI_3$ bilayer in the spin-up FM-like state at an applied field of 6 T. **b,** The magnon frequency in the field range from -7 T to 0 T. The dashed blue line is a linear fit of the acoustic magnon energy indicating a Zeeman shift of $g$ factor ~ 2.1 and intercept of 2.2 $cm^{-1}$ (0.27 meV). Between -0.7 T and -0.75 T, a metamagnetic transition indicated by the grey dashed line occurs as the bilayer switches to the layered AFM state which should host two magnon modes that shift oppositely with applied field. These two magnon modes are illustrated in the bottom-left inset. The green dashed line plots the mode which blueshifts (left cartoon of inset) and is resolvable in our experiments while the orange dashed line indicates the mode which redshifts (right cartoon of inset) into the spectral filter which is indicated by the grey box. The intercept of the green line indicates that the zero-field AFM magnon energy is ~ 3 $cm^{-1}$ (0.37 meV). The error bars represent the uncertainty of the Lorentzian fit used to determine the magnon energy. **c,** Low-frequency Raman spectra taken as the bilayer goes through the transition with increasing magnetic field strength. The switching of the acoustic magnon modes is evident as the peak centered at ~3.6 $cm^{-1}$ at -0.7 T disappears, and a lower energy mode emerges.

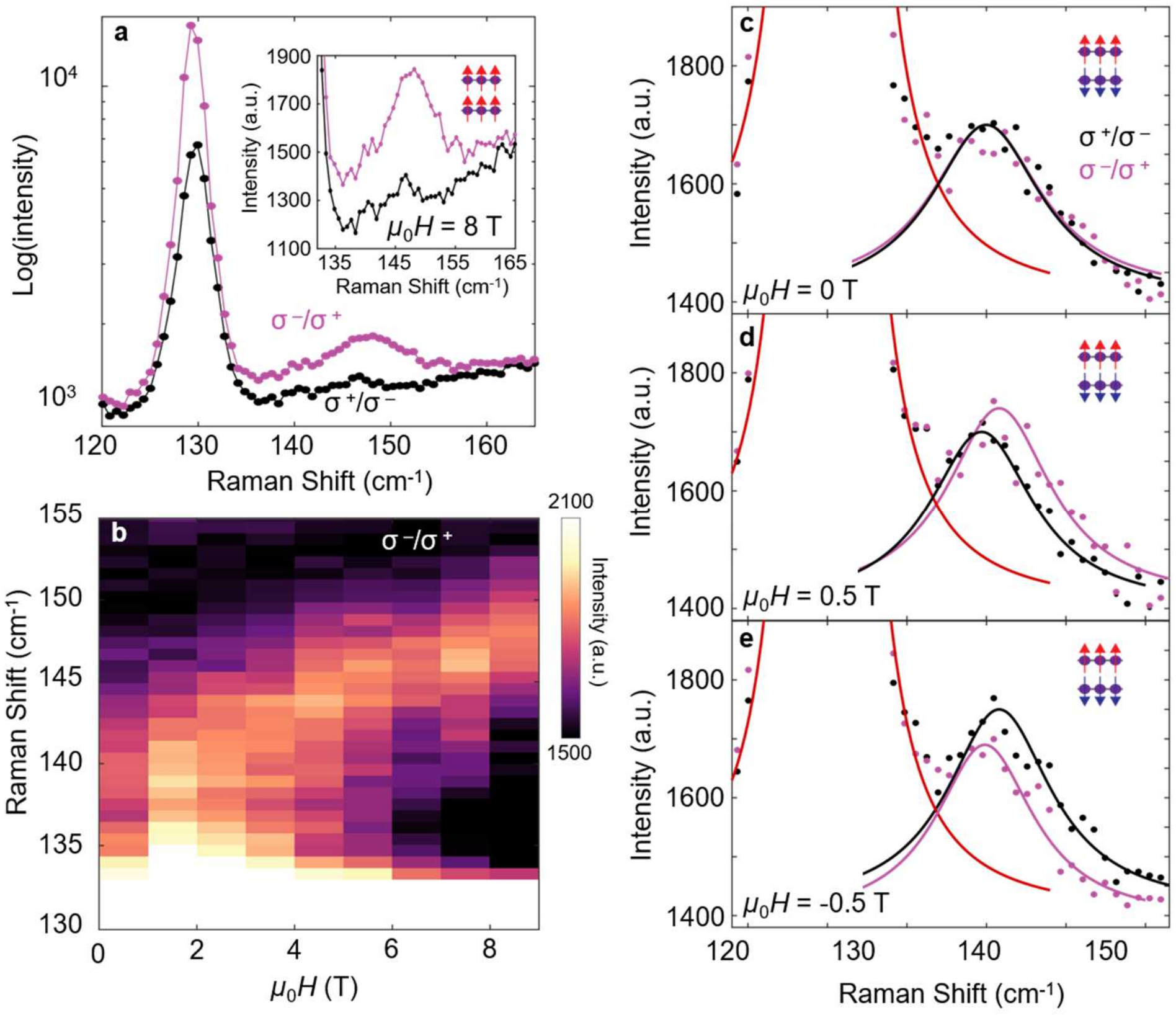

**Figure 4 | Optical magnons in bilayer $CrI_3$. a,** Raman scattering from bilayer $CrI_3$ in an applied field of 8 T with intensity plotted in a log scale. The right inset displays a zoomed in plot of a cross-polarized feature centered at ~148 $cm^{-1}$ (4.4 THz). **b,** Field dependence of the optical magnon from 0 T to 9 T. The feature shifts with a g factor ~ 2, confirming its magnon nature. **c,** Zero-field Raman spectra of bilayer $CrI_3$. The solid lines are Lorentzian functions plotted as guides-to-the-eye. Contrary to the FM-like state which shows a strong favoring of one scattering channel (e.g. $\sigma^-/\sigma^+$ for **a**), the optical magnon in the zero-field AFM state appears equally in both cross-circular channels. **d,** Upon application of magnetic field, there is a splitting between the two optical magnon modes as the $\sigma^-/\sigma^+$ mode blue-shifts and $\sigma^+/\sigma^-$ mode red-shifts. **e,** When the direction of applied field is flipped, the energy shift of the modes is also reversed.

# Supplementary information for

## Direct observation of 2D magnons in atomically thin $CrI_3$

**Authors:** John Cenker[1†], Bevin Huang[1†], Nishchay Suri[2], Pearl Thijssen[1], Aaron Miller[1], Tiancheng Song[1], Takashi Taniguchi[3], Kenji Watanabe[3], Michael A. McGuire[4], Di Xiao[2]*, Xiaodong Xu[1,5]*

[1]Department of Physics, University of Washington, Seattle, Washington 98195, USA
[2]Department of Physics, Carnegie Mellon University, Pittsburgh, Pennsylvania 15213, USA
[3]National Institute for Materials Science, 1-1 Namiki, Tsukuba 305-0044, Japan
[4]Materials Science and Technology Division, Oak Ridge National Laboratory, Oak Ridge, Tennessee 37831, USA
[5]Department of Materials Science and Engineering, University of Washington, Seattle, Washington 98195, USA

[†]These authors contributed equally to this work.

*Corresponding author's e-mail: xuxd@uw.edu; dixiao@cmu.edu

**Content:**

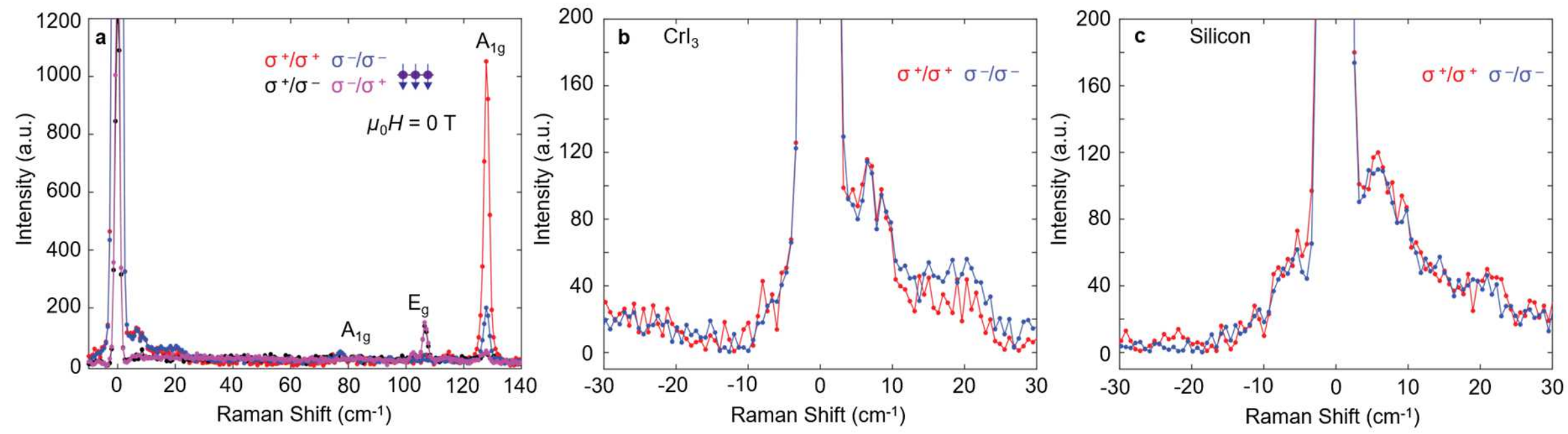

**Figure S1 | Monolayer CrI$_3$ phonon modes and low-frequency co-circularly polarized spectra. a,** Zero-field Raman spectra of monolayer CrI$_3$ showing the standard phonon modes, i.e. $A_{1g}$ phonons at ~127 cm$^{-1}$ and 77 cm$^{-1}$ in the co-circular channels and $E_g$ modes at ~108 cm$^{-1}$ in the cross-circular channel. **b-c,** Low frequency co-circularly polarized Raman scattering on monolayer CrI$_3$ and $SiO_2$/Si substrate. The low-frequency scatter originates from quasi-elastic scattering off the substrate given that the low-frequency co-polarized scatter on monolayer (**b**) is the same as that on silicon (**c**).

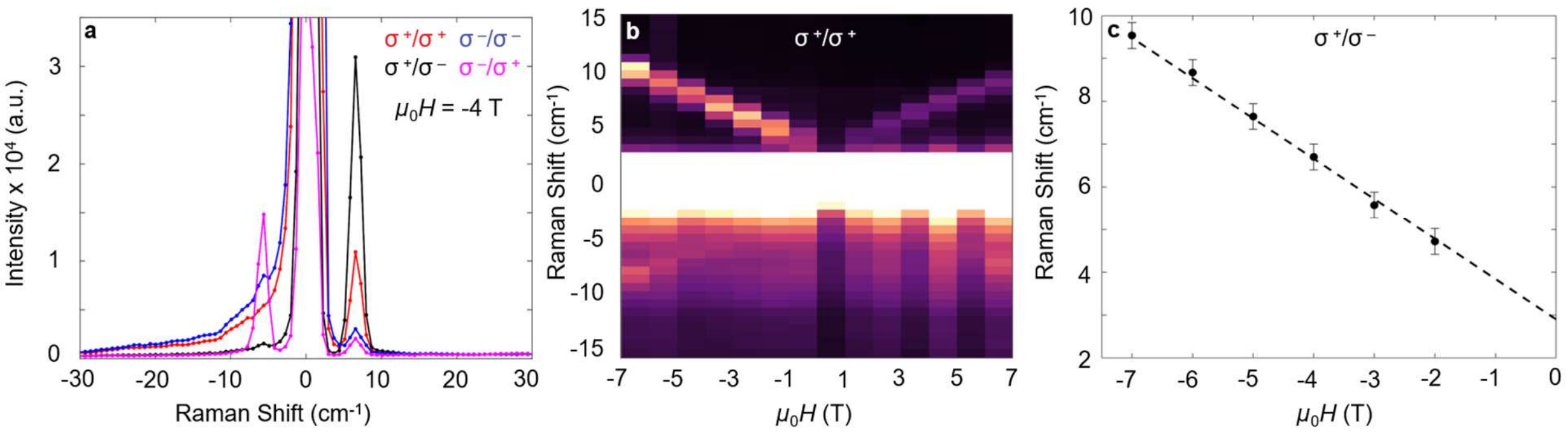

**Figure S2 | Acoustic magnon scattering in unexfoliated single crystal CrI$_3$. a,** Low-frequency circularly polarized Raman spectra of bulk single crystal CrI$_3$. Although the magnon scattering is still predominantly in the cross-circularly polarized channel, a large component appears in the co-circular $\sigma^+/\sigma^+$ channel. **b,** Color map of the magnetic field dependence of $\sigma^+/\sigma^+$ scattering. The linear shift of ~2 $\mu_B$ indicates that it arises from magnon scattering and rules out acoustic phonons. **c,** The energy of the acoustic magnon from the $\sigma^+/\sigma^-$ channel plotted as a function of magnetic field from -7 to -2 T. The dashed line indicates a Zeeman shift of 2 $\mu_B$ with a zero-field energy of ~2.8 cm$^{-1}$ and fits the energy shift of the low-energy peak well. The zero-field energy of the acoustic magnons is determined mainly by the out-of-plane magnetic anisotropy.

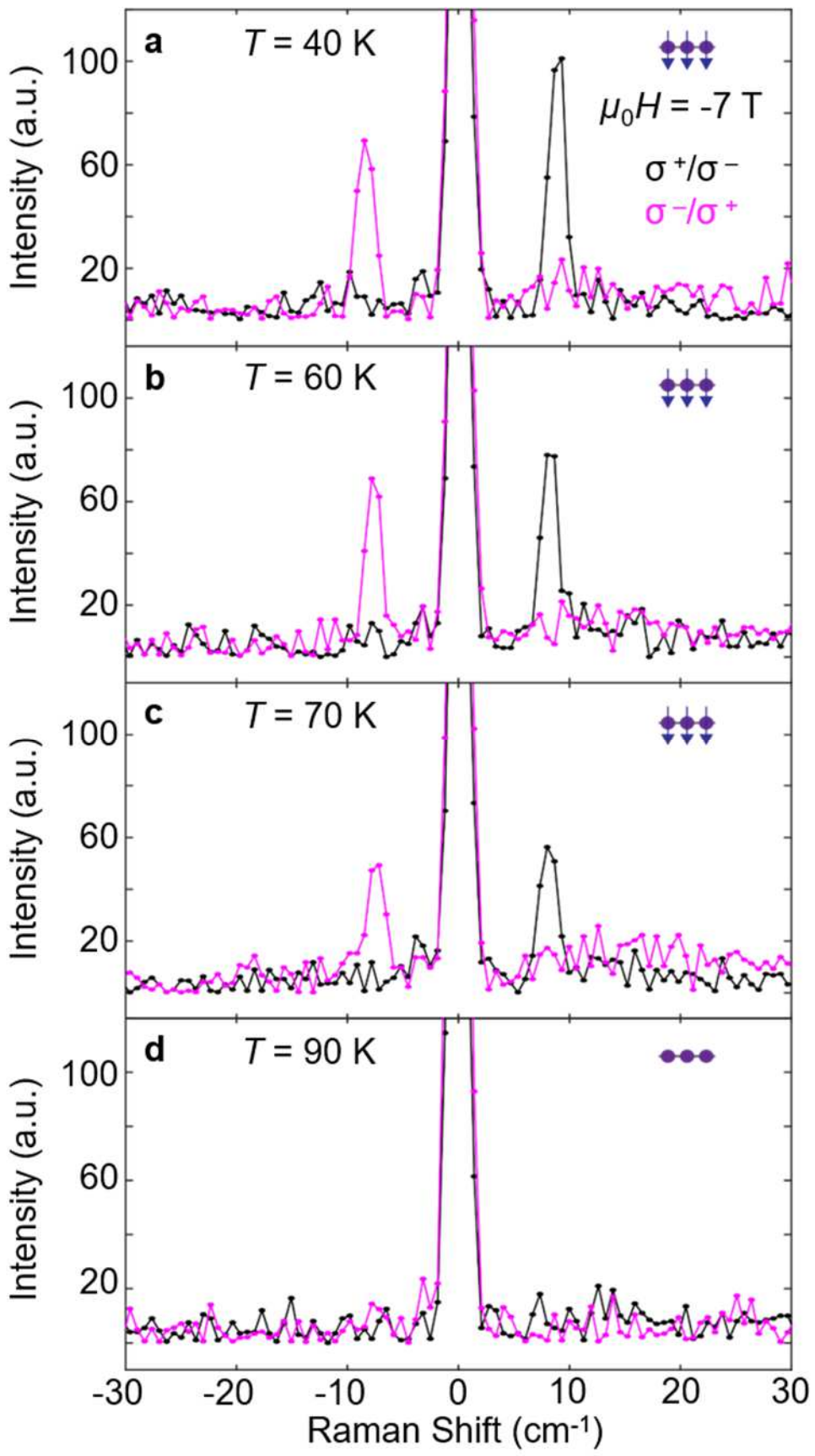

**Figure S3 | Low-frequency Raman spectra of monolayer $CrI_3$ at select temperatures. a-d,** Cross-circularly polarized Raman spectra taken at various temperatures in an applied field of -7 T. Due to the applied field, the magnon is shifted away from the Rayleigh line and is present until ~80 K, well above the Curie temperature of $T_c$ ~ 45 K. Notably, the linewidth does not appear to significantly broaden throughout this temperature range.

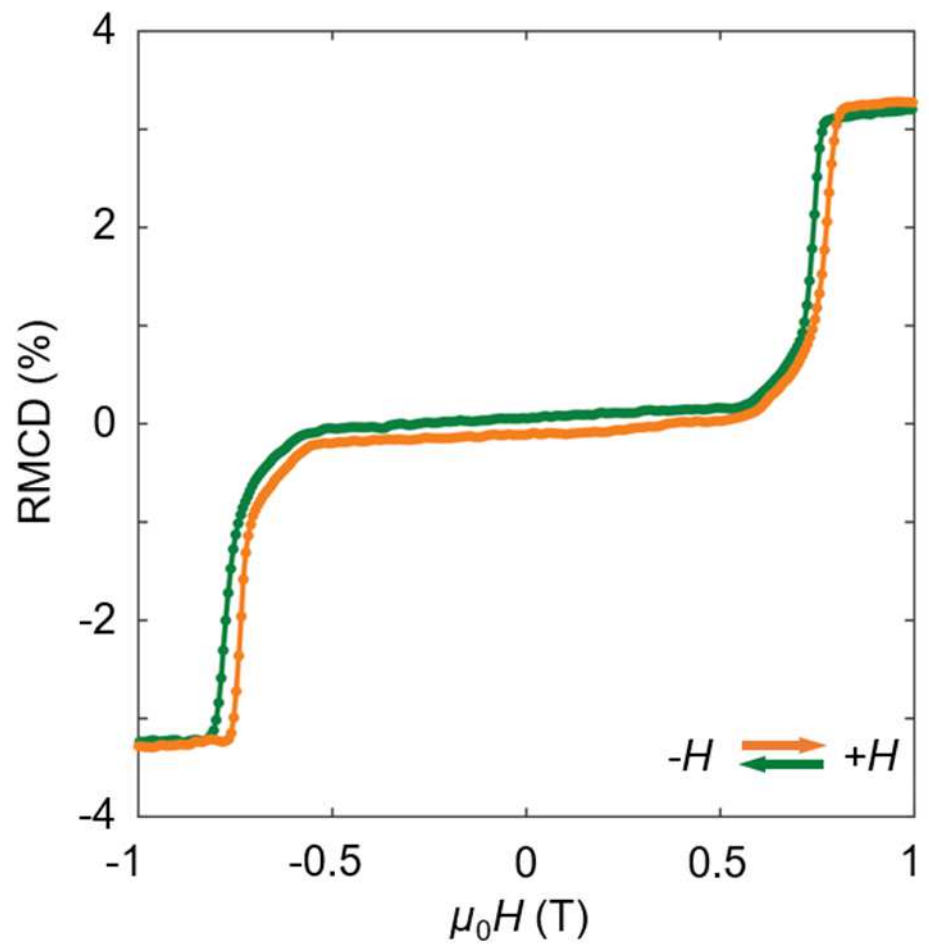

**Figure S4 | RMCD measurement of bilayer $CrI_3$.** Field dependent RMCD measurements of the same bilayer $CrI_3$ flake shown in Fig. 3. The data points in green indicate RMCD measurements taken while the field is swept down while the orange points are taken as the field is swept up. These measurements show the spin-flip transition occurs at around ±0.7 T.

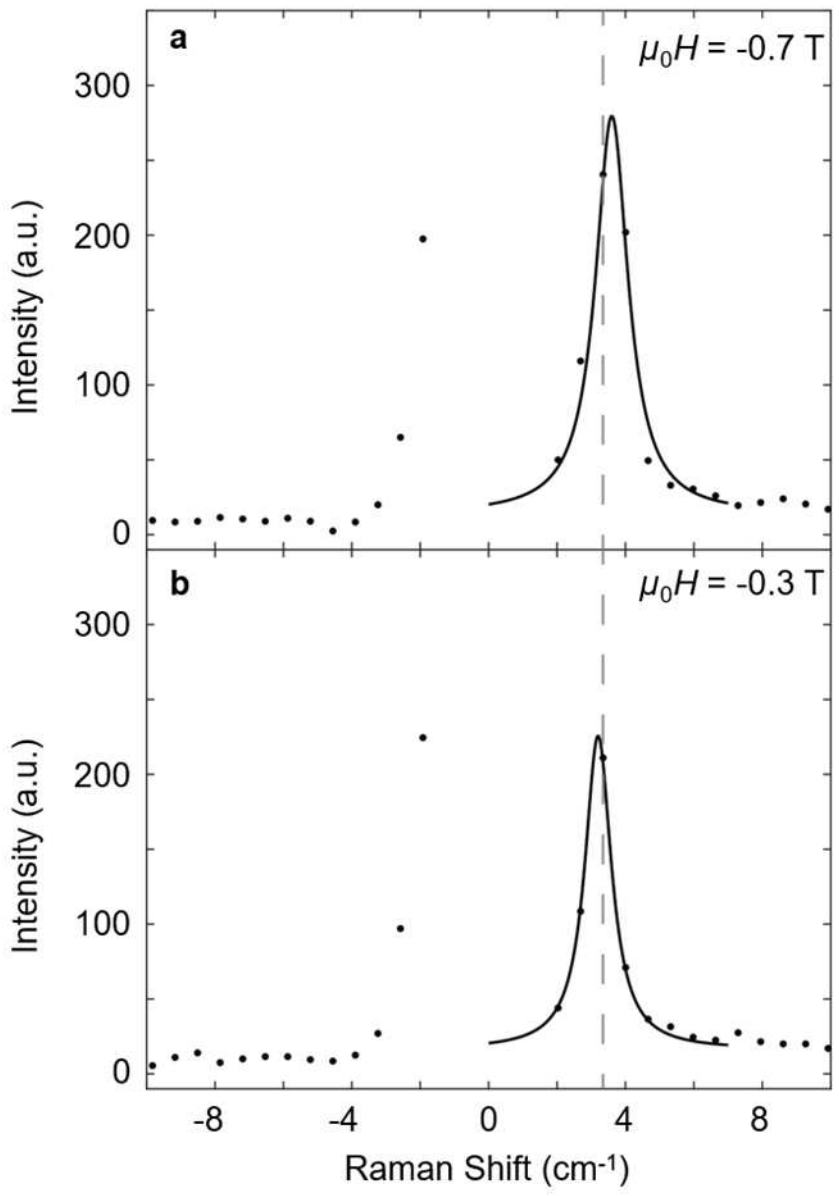

**Figure S5 | Lorentzian fits of the acoustic magnon in AFM bilayer $CrI_3$.** Lorentzian fit of the acoustic magnon in bilayer $CrI_3$ in an applied field of **a,** -0.7 T and **b,** -0.3 T. The center of the Lorentzian fits are at 3.6 $cm^{-1}$ and 3.2 $cm^{-1}$ respectively. The grey dashed line indicates the CCD pixel of the spectrometer with the highest counts of the magnon feature.

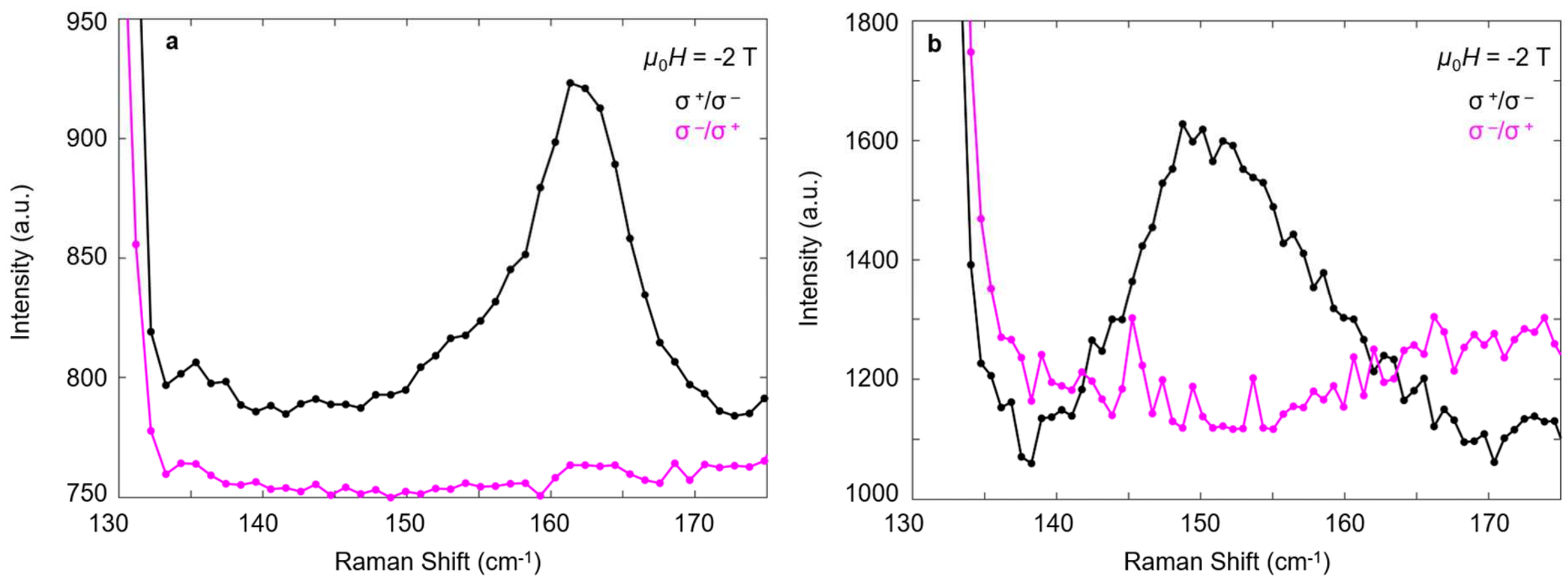

**Figure S6 | Optical magnons in bulk $CrI_3$. a,** Cross-circularly polarized scattering channels in exfoliated thin bulk (~100 nm thickness) showing the optical magnon at an applied field of -2 T. In this sample, the magnon appears at a higher frequency than in bilayer with a zero-field energy of ~160 $cm^{-1}$ (4.8 THz). **b,** Optical magnon in unexfoliated single-crystal $CrI_3$ in an applied field of -2 T. In this thickness, the optical magnon appears at a zero-field frequency of ~148 $cm^{-1}$.